\documentclass[aps,prb,reprint,superscriptaddress]{revtex4-2}
\usepackage[T1]{fontenc}
\usepackage{times}
\usepackage{float}
\usepackage{graphicx,color}
\usepackage{amsfonts,amsmath,amssymb,amsbsy}
\usepackage[colorlinks=true, linkcolor=blue, citecolor=blue, urlcolor=blue]{hyperref}

\def\D{{\mathcal{D}}}

\def\I{{\mathcal{I}}}
\let\mathbf=\boldsymbol

\def\blue#1{\textcolor{blue}{#1}}
\def\emph#1{\textcolor{black}{#1}}
\def\emph#1{\textcolor{magenta}{#1}}
\begin{document}

\title{Janus skyrmion: Interfacial quasiparticle with two-faced helicity}

\author{Xichao Zhang}
\email[Email:~]{xichaozhang@aoni.waseda.jp}
\affiliation{Department of Applied Physics, Waseda University, Okubo, Shinjuku-ku, Tokyo 169-8555, Japan}

\author{Rui Zhang}
\affiliation{Department of Physics, The Hong Kong University of Science and Technology, Clear Water Bay, Kowloon, Hong Kong, China}

\author{Qiming Shao}
\affiliation{Department of Electronic and Computer Engineering, The Hong Kong University of Science and Technology, Clear Water Bay, Kowloon, Hong Kong, China}

\author{Yan Zhou}
\affiliation{Guangdong Basic Research Center of Excellence for Aggregate Science, School of Science and Engineering, The Chinese University of Hong Kong, Shenzhen, Guangdong 518172, China}

\author{\\ Charles Reichhardt}
\affiliation{Theoretical Division and Center for Nonlinear Studies, Los Alamos National Laboratory, Los Alamos, New Mexico 87545, USA}

\author{Cynthia J. O. Reichhardt}
\affiliation{Theoretical Division and Center for Nonlinear Studies, Los Alamos National Laboratory, Los Alamos, New Mexico 87545, USA}

\author{Masahito Mochizuki}
\email[Email:~]{masa_mochizuki@waseda.jp}
\affiliation{Department of Applied Physics, Waseda University, Okubo, Shinjuku-ku, Tokyo 169-8555, Japan}

\begin{abstract}
Janus particles are functional particles with at least two surfaces showing asymmetric properties. We show at the interface between two magnetic regions with different antisymmetric exchange interactions, a new species of two-dimensional topological quasiparticles can emerge, in which different helicity structures can coexist. We name such an interfacial quasiparticle a ``Janus skyrmion'', in analogy to the Janus particle. As the Janus skyrmion shows helicity asymmetry, its size could vary with both the in-plane and out-of-plane magnetic fields. A vertical spin current could drive the Janus skyrmion into one-dimensional motion along the interface without showing the skyrmion Hall effect, at a speed which depends on both the in-plane spin polarization direction and current density. Thermal fluctuations could also lead to one-dimensional random walk of a Brownian Janus skyrmion. This work uncovers unique dynamics intrinsic to interfacial quasiparticles with exotic helicity, which may be realized in interface-engineered magnetic layers.
\end{abstract}

\date{March 5, 2026}


\maketitle

\section{Introduction}
\label{se:Intro}

In chiral magnetic materials, antisymmetric exchange interactions, that is, the Dzyaloshinskii-Moriya (DM) interactions~\cite{Dzyaloshinsky,Moriya}, could give rise to localized spin textures with nontrivial topology~\cite{Mochizuki_JPCM2015,Finocchio_JPD2016,Wiesendanger_NATREVMAT2016,Fert_NATREVMAT2017,Wanjun_PHYSREP2017,Zhang_JPCM2020,Gobel_PHYSREP2021,DelValle_2022,Reichhardt_RMP2022,Ohki_JPCM2024,Liu_RPP2025}, which are usually treated as quasiparticles as they have well-defined shape, nanoscale size, and good mobility~\cite{Reichhardt_RMP2022}.
These magnetic quasiparticles could be dynamically manipulated by external drives, such as spin currents~\cite{Reichhardt_RMP2022,Zang_PRL2011,Wanjun_SCIENCE2015,Wanjun_NPHYS2017,Litzius_NPHYS2017,Tomasello_SREP2014,Zhang_Laminar2023,Raab_PRE2024,Sampaio_NN2013,Zhang_PNAS2025,Lee_PRB2025}, and are considered to be next-generation information carriers for both conventional and quantum devices~\cite{Kang_PIEEE2016,Li_MH2021,Luo_APLM2021,Marrows_APL2021,Lohani_PRX2019,Psaroudaki_PRL2021,Psaroudaki_PRB2022,Xia_PRL2022,Hou_2020,Hou_2022,Hou_2025}.
Although different magnetic quasiparticles may carry identical integer topological charge~\cite{Zhang_JPCM2020,Ohki_JPCM2024}, which is described by $Q=\frac{1}{4\pi}\int\boldsymbol{m}\cdot(\frac{\partial\boldsymbol{m}}{\partial x}\times\frac{\partial\boldsymbol{m}}{\partial y})dxdy$ with $\boldsymbol{m}$ being the reduced magnetization, their static and dynamic properties depend on their internal structures~\cite{Zhang_JPCM2020,Ohki_JPCM2024}.

\begin{figure*}[t]
\centerline{\includegraphics[width=1.00\textwidth]{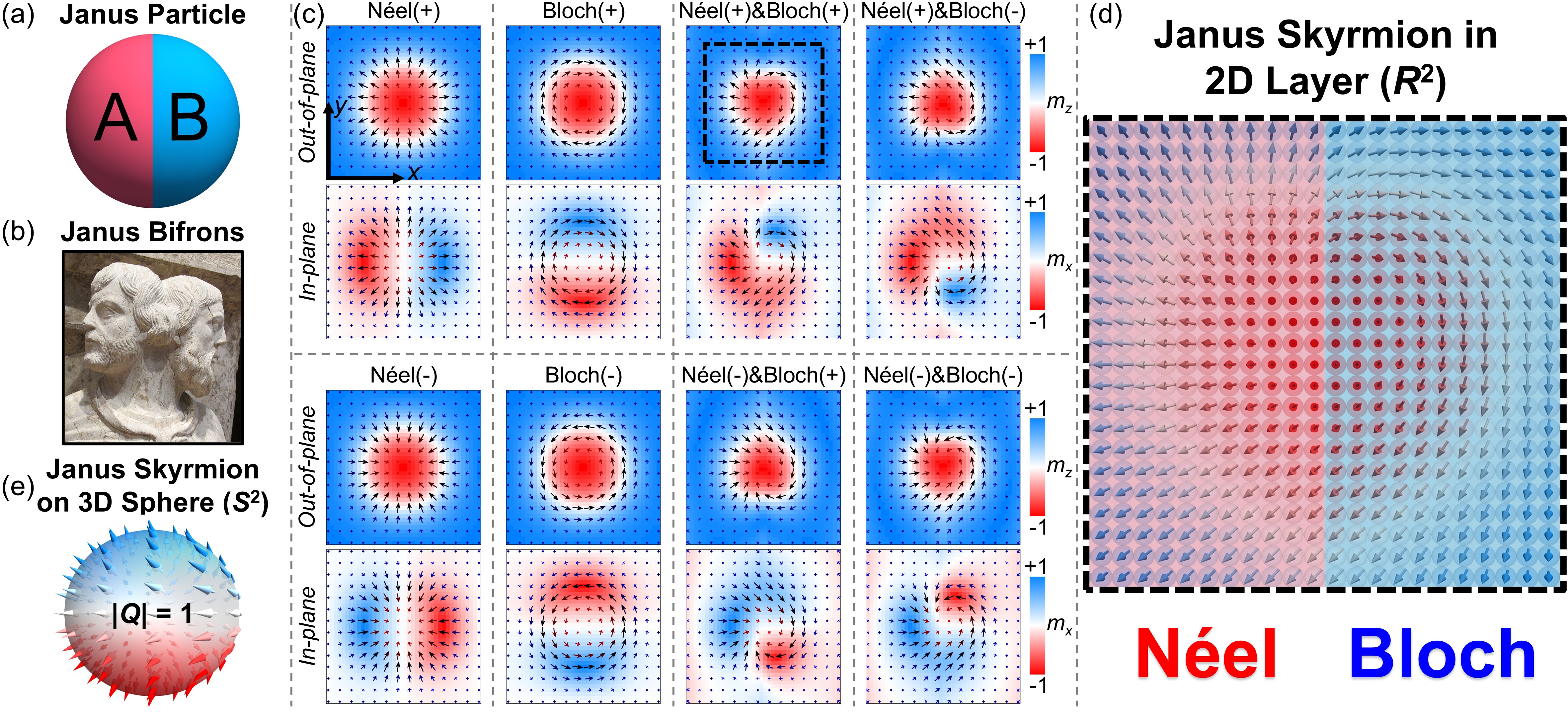}}
\caption{%
Janus skyrmion quasiparticle with two-faced asymmetric helicity at the interface.
(a) A basic Janus particle has two distinct surfaces, that is, the left side A and right side B, which show different or asymmetric properties.
(b) The statue of the two-faced god Janus Bifrons, the ancient Roman god of beginnings and transitions, who is able to look at opposite directions. The image is a cropped version from wikipedia.org, licensed under the \href{https://creativecommons.org/licenses/by-sa/3.0}{Creative Commons Attribution-Share Alike 3.0 Unported License}.
(c) Top view of various relaxed N{\'e}el-type, Bloch-type, and Janus skyrmions with $Q=-1$. Skyrmions are fully relaxed at the center of a square magnetic layer with the same or two different types of DM interactions.
The length, width, and thickness of the monolayer are equal to $40$ nm, $40$ nm, and $1$ nm, respectively. The cell size is $1\times 1\times 1$ nm$^{3}$. No external magnetic field is applied.
For N{\'e}el-type and Bloch-type skyrmions, the DM interaction is uniform and of the same type in the model. For Janus skyrmions, the value of N{\'e}el-type DM interaction in the left region ($x=0-20$ nm) is fixed at $D_{\text{L}}=\pm 3.7$ mJ m$^{-2}$, and the strength of Bloch-type DM interaction in the right region ($x=20-40$ nm) is set to $|D_{\text{R}}|=|D_{\text{L}}|$. The type and sign of $D_{\text{L}}$ and $D_{\text{R}}$ are indicated beyond each skyrmion configuration.
A minimum value of $|D_{\text{R}}|=|D_{\text{L}}|=3.1$ mJ m$^{-2}$ is required to stabilize the Janus skyrmion.
The color scale represents the reduced out-of-plane ($m_z$) or in-plane ($m_x$) spin component. The spin configurations are indicated by black arrows.
(d) A basic Janus skyrmion has two distinct helicity structures. For example, the left half is with N{\'e}el-type helicity, and the right half is with Bloch-type helicity. The shape of the Janus skyrmion is thus like a heart, which is in stark contrast to a conventional skyrmion with centrosymmetric helicity.
(e) Topological mapping of the spin configurations of a 2D Janus skyrmion with $Q=-1$ [(d)] onto a three-dimensional (3D) $2$-sphere.
}
\label{FIG1}
\end{figure*}

In principle, the size, shape, and internal structures of topological spin textures can be modified by external means~\cite{Liu_RPP2025}.
An important internal degree of freedom of particle-like topological spin textures is their helicity~\cite{Zhang_JPCM2020,Ohki_JPCM2024,Zhang_NC2017,Zhang_JMSJ2023,Wang_AM2025,Dai_SA2023,Srivastava_NL2018,Han_NL2025,Yao_NJP2020,Kuchkin_PRB2020}.
For example, among the magnetic quasiparticle family, the skyrmion is an exemplary member that can be found in ultra-thin magnetic layers and low-dimensional magnets~\cite{Roszler_NATURE2006,Muhlbauer_SCIENCE2009,Yu_NATURE2010,Mochizuki_JPCM2015,Finocchio_JPD2016,Wiesendanger_NATREVMAT2016,Fert_NATREVMAT2017,Wanjun_PHYSREP2017,Zhang_JPCM2020,Gobel_PHYSREP2021,DelValle_2022,Reichhardt_RMP2022,Ohki_JPCM2024,Liu_RPP2025}, of which the helicity is mainly determined by the type of DM interactions~\cite{Zhang_JPCM2020,Tomasello_SREP2014}.
It shows N{\'e}el-type helicity in magnetic layers with interface-induced DM interactions~\cite{Wanjun_SCIENCE2015,Wanjun_NPHYS2017,Litzius_NPHYS2017}, and shows Bloch-type helicity in magnets with bulk DM interactions (Fig.~\ref{FIG1})~\cite{Roszler_NATURE2006,Muhlbauer_SCIENCE2009,Yu_NATURE2010}.
For either case, the helicity structure of a two-dimensional (2D) skyrmion is centrosymmetric with respect to the skyrmion center.
The 2D skyrmions with helicity asymmetry as well as their static and dynamic properties remain insufficiently explored.

In nanoscience, Janus particles [Fig.~\ref{FIG1}(a)] are a special type of highly functional particles named after the god Janus [Fig.~\ref{FIG1}(b)], which have at least two surfaces showing different or asymmetric physical properties~\cite{Walther_SM2008,Walther_CR2013}.
They can be fabricated at nanoscale and microscale dimensions, and due to their multifunctionalized surface structures, they play important roles in various fields, including soft matter, active matter, biomedical science, and electronics~\cite{Walther_SM2008,Walther_CR2013,Li_AC2011,Zhang_L2017,Su_MTB2019,Zhang_NRM2021}.
For example, catalytically active Janus particles with asymmetric reaction surfaces can demonstrate self-propulsion and thus, can be used for active matter experiments~\cite{Paxton_JACS2004}.
Also, magnetoactive Janus particles can be used to build a multifunctional swarm metamaterial system for programmable dynamic display, nonvolatile memory, and information encryption~\cite{Zhang_AM2024}.

The concept of Janus particles could be shifted to magnetic quasiparticles, which have at least two different or asymmetric internal physical structures. For example, 2D skyrmions with asymmetric helicity structures may be treated as Janus quasiparticles in magnetic materials.
Most recently, experimentalists have demonstrated the possibility to engineer the sign of the DM interaction~\cite{Niu_NC2025,Niu_NC2024}, which is an effective way to induce helicity asymmetry in 2D skyrmions and will be the key to realize Janus skyrmions.

In this work, we show that one may artificially construct a Janus skyrmion with unique helicity asymmetry at the engineered interface between two magnetic regions with different types and signs of DM interactions. We explore the dynamics of such a Janus skyrmion driven by magnetic fields, spin currents, or thermal fluctuations. We find that the Janus skyrmion can be viewed as a hybridization of N{\'e}el-type and Bloch-type skyrmions. However, its mobility is limited by the one-dimensional (1D) interface, leading to exotic properties that cannot be found in conventional N{\'e}el-type and Bloch-type skyrmions.

\begin{figure}[t]
\centerline{\includegraphics[width=0.49\textwidth]{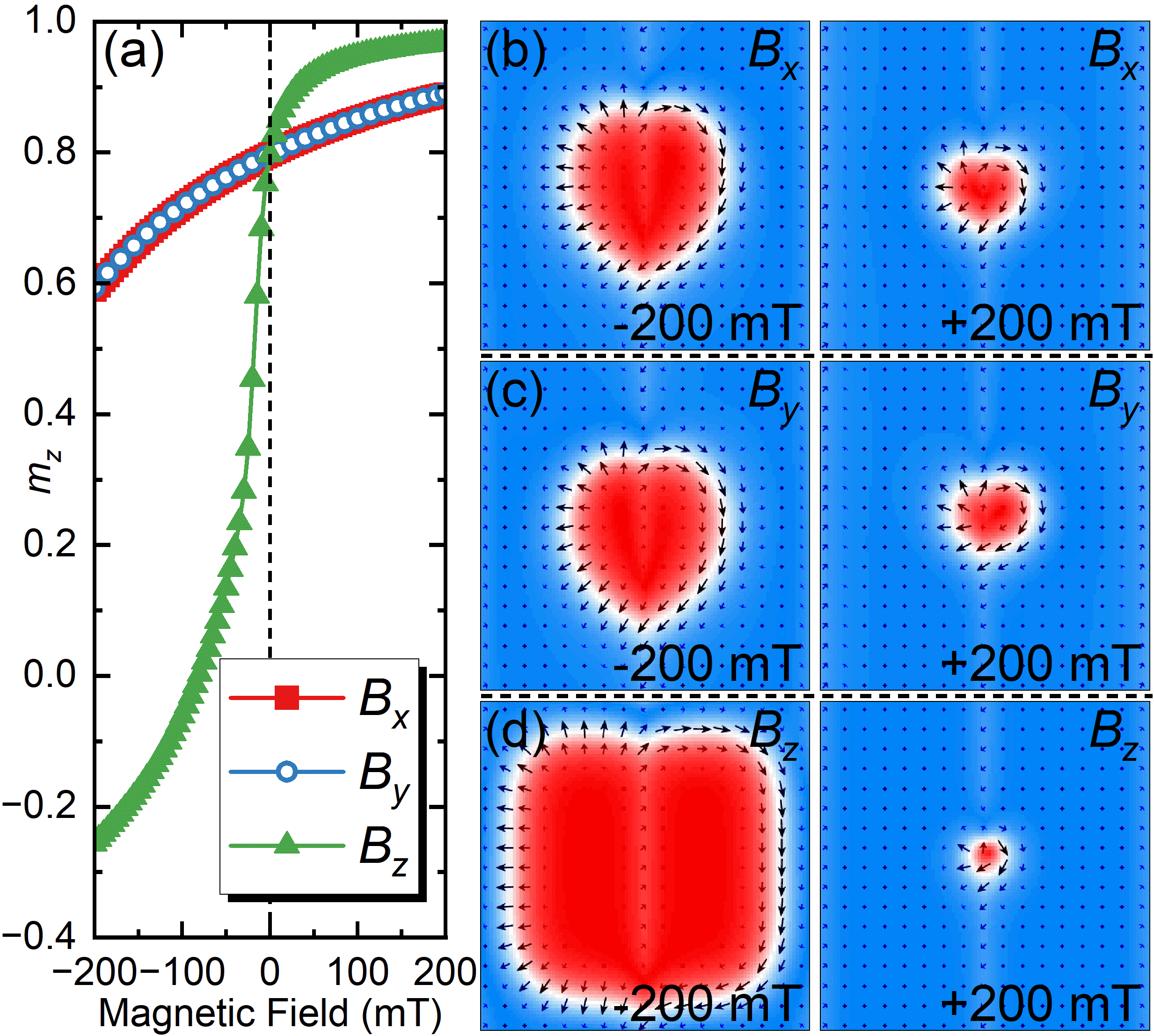}}
\caption{%
Manipulation of a static Janus skyrmion by in-plane or out-of-plane magnetic field.
(a) Reduced out-of-plane magnetization ($m_z$) of the model as functions of applied external magnetic fields. The magnetic field is applied in the $x$, $y$, or $z$ direction, which scans from $-200$ mT to $+200$ mT.
(b) Snapshots of the Janus skyrmion at $B_x=\pm 200$ mT.
(c) Snapshots of the Janus skyrmion at $B_y=\pm 200$ mT.
(d) Snapshots of the Janus skyrmion at $B_z=\pm 200$ mT.
Here, the Janus skyrmion is stabilized at the interface between N{\'e}el-type (left) and Bloch-type (right) DM interactions. $D_{\text{L}}=D_{\text{R}}=3.5$ mJ m$^{-2}$. The model size is $100\times 100\times 1$ nm$^{3}$. Other parameters are given in Fig.~\ref{FIG1} caption.
}
\label{FIG2}
\end{figure}

\section{Results and Discussion}
\label{se:Results}

\subsection{From Janus particle to Janus skyrmion}
\label{se:Janus}

We first study the static structure of a Janus skyrmion.
A Janus particle has two faces with different properties [Fig.~\ref{FIG1}(a)], therefore, we focus on a Janus skyrmion with two different helicity structures, as illustrated in Figs.~\ref{FIG1}(c)~and~\ref{FIG1}(d). The left half of the Janus skyrmion is of N{\'e}el-type helicity, while the right half is of Bloch-type helicity.
Such a hybrid helicity structure is not centrosymmetric with respect to the skyrmion center, which is different to skyrmions with intermediate helicity between N{\'e}el-type and Bloch-type ones~\cite{Kim2018,Kong2024,Liu2024,Akhir2024,Dai_SA2023}.
Consequently, the Janus skyrmion with helicity asymmetry carries a topological charge of $|Q|=1$ but has a unique heart-like shape.
Previously, heart-like skyrmions are also found in chiral skyrmions with chiral kinks~\cite{Kuchkin_PRB2020} as well as in modified multilayer magnets~\cite{Rare_Metals_2022}.

We assume that a Janus skyrmion with $Q=-1$ and two different helicity structures is formed at the interface between two magnetic regions with different types of DM interactions.
In our simulation [Fig.~\ref{FIG1}(c)], we place and relax a skyrmion at the center of a magnetic layer. The size of the magnetic layer is $40\times 40\times 1$ nm$^{3}$, which includes $1,600$ spins.
The spin dynamics is governed by the Landau-Lifshitz-Gilbert (LLG) equation $\partial_{t}\boldsymbol{m}=-\gamma_{0}\boldsymbol{m}\times\boldsymbol{h}_{\text{eff}}+\alpha(\boldsymbol{m}\times\partial_{t}\boldsymbol{m})$, which is solved by using the OOMMF simulator~\cite{OOMMF}.
In the LLG equation, the absolute gyromagnetic ratio $\gamma_{0}=2.211\times 10^{5}$ m A$^{-1}$ s$^{-1}$ and the damping parameter $\alpha=0.3$ (for dynamics simulation).
The effective field $\boldsymbol{h}_{\text{eff}}=-\frac{1}{\mu_{0}M_{\text{S}}}\cdot\frac{\delta\varepsilon}{\delta\boldsymbol{m}}$, where $\mu_{0}$ and $\varepsilon$ denote the vacuum permeability constant and average energy density, respectively.
The average energy density includes the contributions from the ferromagnetic exchange interaction, DM interactions (N{\'e}el-type and Bloch-type), perpendicular magnetic anisotropy (PMA), demagnetization, and applied magnetic field, given as
$\varepsilon=A\left(\nabla\boldsymbol{m}\right)^{2}+D_{\text{N}}\left[m_{z}\left(\boldsymbol{m}\cdot\nabla\right)-\left(\nabla\cdot\boldsymbol{m}\right)m_{z}\right]+D_{\text{B}}\left[\boldsymbol{m}\cdot\left(\nabla\times\boldsymbol{m}\right)\right]-K(\boldsymbol{n}\cdot\boldsymbol{m})^2-\frac{M_{\text{S}}}{2}(\boldsymbol{m}\cdot\boldsymbol{B}_{\text{d}})-M_{\text{S}}(\boldsymbol{m}\cdot\boldsymbol{B}_{\text{a}})$.
Note that $D_{\text{N}}$ and $D_{\text{B}}$ stand for the N{\'e}el-type (i.e., interfacial-like) and Bloch-type (i.e., bulk-like) DM interaction parameters, respectively.
$\boldsymbol{B}_{\text{d}}$ is the demagnetization field, $\boldsymbol{B}_{\text{a}}$ is the applied magnetic field, and $\boldsymbol{n}$ is the unit surface normal vector.
The default magnetic parameters are~\cite{Tomasello_SREP2014,Zhang_Laminar2023,Sampaio_NN2013,Xichao_NC2016,Xichao_PRB2016A,Xichao_PRB2016B}: the saturation magnetization $M_{\text{S}}=580$ kA m$^{-1}$, the exchange constant $A=15$ pJ m$^{-1}$, and PMA constant $K=0.8$ MJ m$^{-3}$.
These magnetic parameters are based on an as-deposited ultra-thin magnetic film (e.g., cobalt) on a heavy-metal layer (e.g., platinum) grown by sputtering deposition~\cite{Sampaio_NN2013}, which can effectively stabilize nanoscale skyrmions.
For modeling purposes, we consider different types of DM interactions in the left and right regions [Fig.~\ref{FIG1}(d)].
The interfacial-like DM interaction in the left half of the model is of the $C_{nv}$ symmetry group~\cite{Rohart_PRB2013} and the corresponding DM interaction vectors ($\vec{d}_{ij}$) are given in a counterclockwise configuration (i.e., $\vec{d}_{ij}=\vec{u}_{ij}\times\hat{z}$ with $\vec{u}_{ij}$ being the unit vector between site $i$ and $j$, and $\hat{z}$ being the surface normal).
In the right half of the model, the bulk-like DM interaction is of the $D_{n}$ symmetry group~\cite{Niu_NC2024} and the corresponding DM vectors are given in a radical configuration (i.e., $\vec{d}_{ij}=\vec{u}_{ij}$).
In principle, the configuration of DM interaction vectors can be modified by controlling symmetry, spin-orbit coupling, and local atomic environment at the interface between the magnetic layer and heavy-metal substrate.
In experiments, these modifications can be realized by applying an electric field on a local region, inducing strain locally, or changing the interface composition or stacking order.
Particularly, to change the DM interaction vector configuration from interfacial-like to bulk-like, one can apply an electric field~\cite{Dai_SA2023,Srivastava_NL2018,Han_NL2025,Yao_NJP2020,Hou_2020,Hou_2022} only to the right half of the sample in a partially voltage-gated device structure.
A recent experimental report on voltage-controlled DM interactions has demonstrated the possibility to deform N{\'e}el-type skyrmions into Bloch-type ones by applying an electric field to control the DM interaction vectors~\cite{Dai_SA2023}.
The Janus skyrmions could be realized at the boundary between voltage-controlled and uncontrolled regions.
Also, a previous study suggested that one can construct a heart-shaped Janus skyrmion by utilizing the edge effect of the oxide layer upon a skyrmion-hosting magnetic layer~\cite{Rare_Metals_2022}.

As shown in Fig.~\ref{FIG1}(c), if we consider uniform N{\'e}el-type or Bloch-type DM interaction in the magnetic layer (either positive or negative), the relaxed skyrmion is of N{\'e}el-type or Bloch-type, where the helicity structure is centrosymmetric with respect to the skyrmion center.
However, if we assume a N{\'e}el-type DM interaction in the left half of the magnetic layer ($x\in[0~\text{nm}, 20~\text{nm}]$), and a Bloch-type DM interaction in the right half of the magnetic layer ($x\in[20~\text{nm}, 40~\text{nm}]$), the relaxed skyrmion at the interface between the two different magnetic regions shows hybrid and asymmetric helicity structures, including both N{\'e}el-type and Bloch-type spin configurations.
Such a Janus skyrmion is stable at the $D_{\text{N}}$-$D_{\text{B}}$ interface and its topological charge is $Q=-1$ [Fig.~\ref{FIG1}(e)].
Its asymmetric internal spin configurations can be modified by changing the signs of $D_{\text{N}}$ and $D_{\text{B}}$ [Fig.~\ref{FIG1}(c)].
Note that the Janus skyrmion stabilized by positive $D_{\text{N}}$ and $D_{\text{B}}$ could be transformed to that stabilized by negative $D_{\text{N}}$ and $D_{\text{B}}$ through the symmetry operation ($m_x$, $m_y$, $m_z$) $\rightarrow$ ($-m_x$, $-m_y$, $m_z$).
In Supplementary Fig.~\blue{1}~\cite{SM}, we show that the Janus skyrmion also exhibits non-uniform distributions of both topological charge density and total energy density, which are in sharp contrast to that of conventional N{\'e}el-type and Bloch-type skyrmions.
Moreover, the size and shape of the Janus skyrmion depend on the DM interaction strength in each region. It may be deformed when the DM interactions in the left and right regions are not identical (Supplementary Fig.~\blue{2}~\cite{SM}).

\subsection{In-plane and out-of-plane magnetic fields}
\label{se:Field}

In Fig.~\ref{FIG2}, we apply an external magnetic field $\boldsymbol{B}_{\text{a}}$ to a Janus skyrmion, and observe its response to the field.
The Janus skyrmion is initially relaxed in a magnetic layer of $100\times 100\times 1$ nm$^{3}$, which includes $10,000$ spins. The DM interaction in the left half of the magnetic layer is of N{\'e}el-type, and it is of Bloch-type in the right half. The relaxed Janus skyrmion at zero magnetic field shows asymmetric helicity, as given in Fig.~\ref{FIG1}(d).

In theory, a conventional N{\'e}el-type or Bloch-type skyrmion [Fig.~\ref{FIG1}(c)] cannot be manipulated by an in-plane magnetic field, as the magnetic field forces acting on the in-plane spins of the skyrmion will cancel with each other due to the centrosymmetry of the helicity structure. Only a strong in-plane magnetic field may deform the skyrmion into a noncircular shape by dragging perpendicularly magnetized spins into the $x$-$y$ plane~\cite{Lin_PRB2015}.
However, we find that the size of the Janus skyrmion with helicity asymmetry can be adjusted by an in-plane magnetic field. In Fig.~\ref{FIG2}, an external magnetic field applied in the $-x$ or $-y$ direction could increase the skyrmion size, while a field applied in the $+x$ or $+y$ direction could shrink the skyrmion.
This is a unique feature that cannot be found in conventional circular skyrmions with centrosymmetric helicity structures.
Indeed, similar to the case of a conventional skyrmion, an out-of-plane magnetic field applied in the $\pm z$ direction can modify the size of the Janus skyrmion, which is an expected result.
We note that the light stripe along the interface between the left and right regions, visible in Fig.~\ref{FIG2}, is a result of the difference between the DM interactions in the left and right regions. The DM interactions lead to micromagnetic boundary conditions that tilt boundary spins at the interface~\cite{Rohart_PRB2013}. Namely, the interfacial-like DM interaction ensures that the boundary spins rotate in a plane containing the boundary surface normal, while the bulk-like DM interaction ensures that the boundary spins rotate in a plane parallel to the boundary surface~\cite{Rohart_PRB2013}. Consequently, due to the uniform symmetric ferromagnetic exchange coupling in the model, the spins exhibit a smooth variation across the interface in the perpendicularly magnetized layer, while a tiny in-plane spin component arises from the different tilt directions of the spins adjacent to the interface.

\begin{figure}[t]
\centerline{\includegraphics[width=0.50\textwidth]{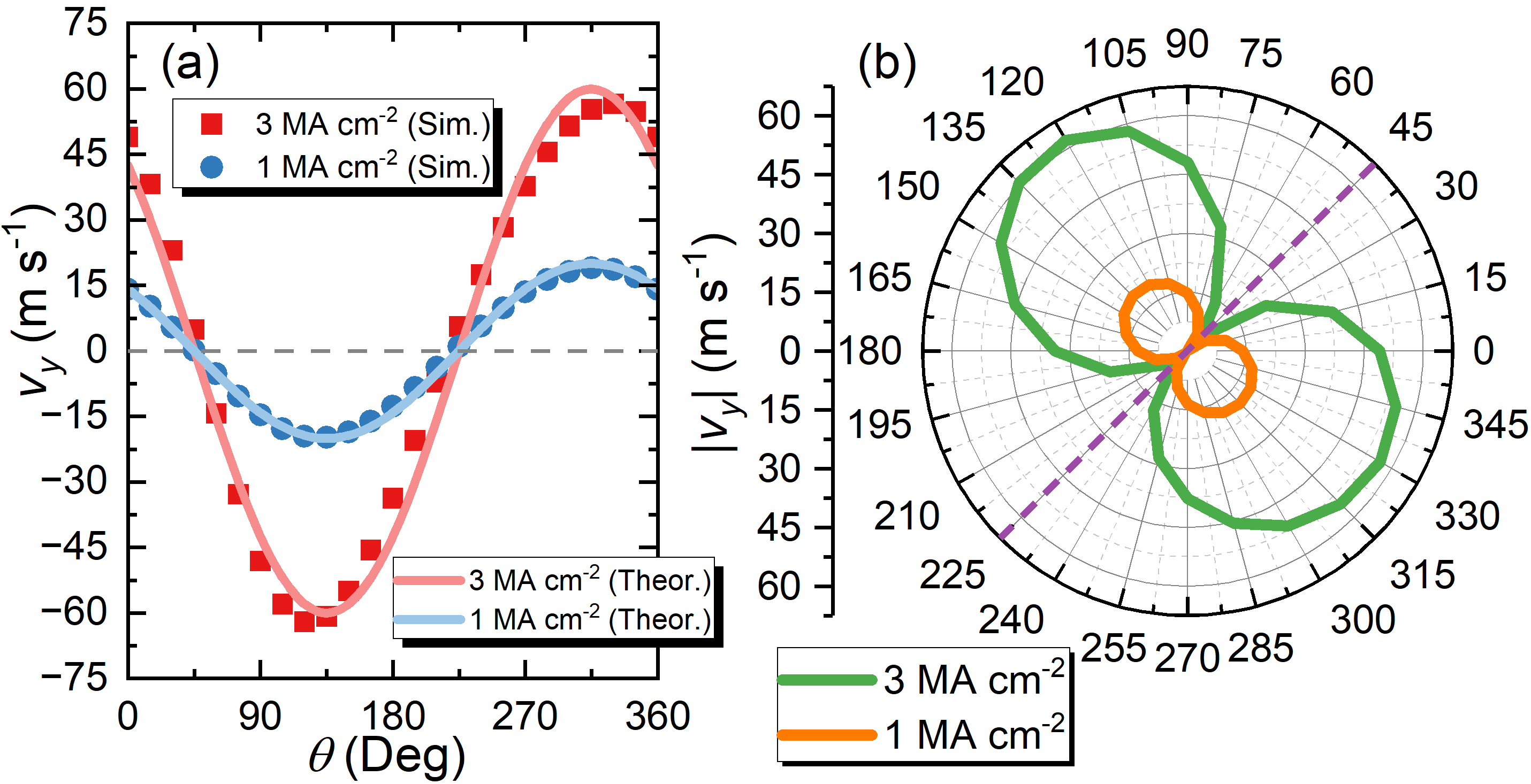}}
\caption{%
Current-induced dynamics of a Janus skyrmion at the interface ($\theta$ dependence).
(a) The velocity of the Janus skyrmion along the interface (i.e., $\pm y$ directions) as functions of the spin polarization angle ($\theta$). Symbols are simulation results, and curves show theoretical solutions.
(b) The $\theta$-dependent speed of the Janus skyrmion moving along the interface. The current density $j$ is fixed at $1$ or $3$ MA cm$^{-2}$. Only simulation results are shown.
Here, the Janus skyrmion is stabilized at the interface between N{\'e}el-type (left) and Bloch-type (right) DM interactions. $D_{\text{L}}=D_{\text{R}}=3.5$ mJ m$^{-2}$. The model size is $100\times 100\times 1$ nm$^{3}$. Other parameters are given in Fig.~\ref{FIG1} caption.
}
\label{FIG3}
\end{figure}

\begin{figure}[t]
\centerline{\includegraphics[width=0.50\textwidth]{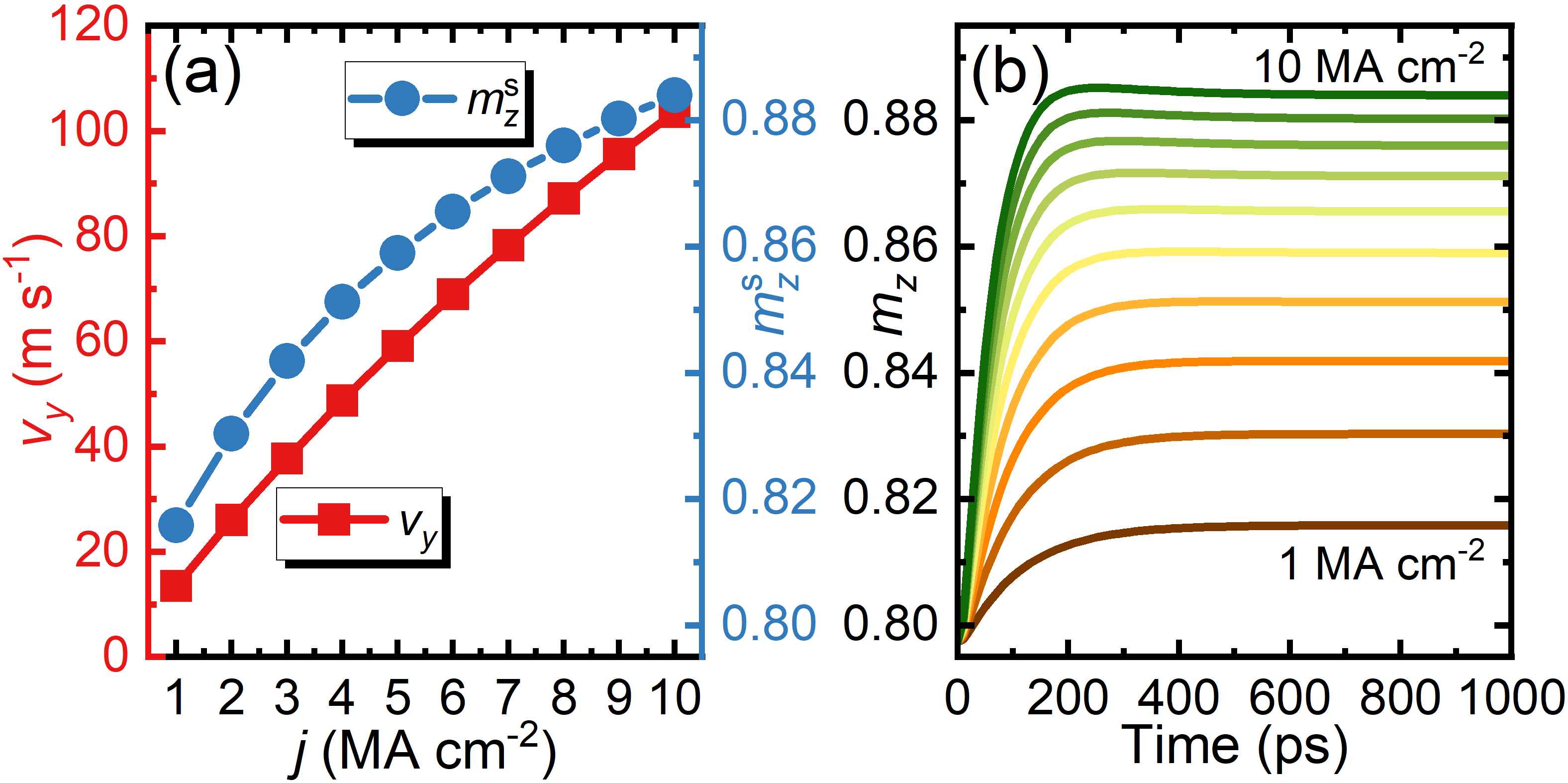}}
\caption{%
Current-induced dynamics of a Janus skyrmion at the interface ($j$ dependence).
(a) Left: The velocity of the Janus skyrmion along the interface (i.e., $\pm y$ directions) as a function of the current density ($j$). Right: The reduced out-of-plane magnetization of the system obtained during the stable motion ($m_z^s$) as a function of $j$. The spin polarization angle is fixed at $\theta=270$ degrees (i.e., $\boldsymbol{p}=-\hat{y}$).
(b) The time-dependent $m_z$ during the motion of a Janus skyrmion driven by different current densities ($j=1-10$ MA cm$^{-2}$).
See Fig.~\ref{FIG3} caption for other parameters.
}
\label{FIG4}
\end{figure}

\subsection{Current-induced dynamics}
\label{se:Current}

The Janus skyrmion is an interfacial state stabilized at the interface. Although it is a 2D topological spin texture, its dynamics will be affected by the interface and should be largely different to that of a conventional skyrmion. Here, we explore the dynamics of a Janus skyrmion at the interface driven by a vertical spin current~\cite{Wanjun_SCIENCE2015,Wanjun_NPHYS2017,Litzius_NPHYS2017}.
Initially, a Janus skyrmion is relaxed at the center of a magnetic layer ($100\times 100\times 1$ nm$^{3}$) with periodic boundary conditions applied in both $x$ and $y$ directions. The DM interaction in the left half of the magnetic layer is of N{\'e}el-type, and it is of Bloch-type in the right half.
We then apply a vertical spin current to drive the Janus skyrmion into motion, which can be realized by the spin Hall effect in the substrate underneath the magnetic layer~\cite{Wanjun_SCIENCE2015,Wanjun_NPHYS2017,Litzius_NPHYS2017,Tomasello_SREP2014,Zhang_Laminar2023}.

To simulate current-induced dynamics, a damping-like spin Hall torque is included in the right-hand side of the LLG equation, given as
$\boldsymbol{\tau}_{\text{d}}=u\left(\boldsymbol{m}\times\boldsymbol{p}\times\boldsymbol{m}\right)$.
The coefficient $u=\left|\left(\gamma_{0}\hbar/\mu_{0}e\right)\right|\cdot\left(j\theta_{\text{SH}}/2aM_{\text{S}}\right)$,
where $\hbar$ is the reduced Planck constant, $e$ is the electron charge, $a$ is the magnetic layer thickness, $j$ is the current density, $\theta_{\text{SH}}=1$ is the spin Hall angle, and $\boldsymbol{p}$ is the spin polarization direction.
The spin-polarization angle between $\boldsymbol{p}$ and the $+x$ direction is defined as $\theta$, which can be adjusted by tuning the in-plane electron flow direction in the substrate~\cite{Tomasello_SREP2014,Zhang_Laminar2023}.

In stark contrast to a conventional skyrmion driven by the spin Hall torque, of which the velocity direction (i.e., the skyrmion Hall angle) depends on $\theta$ but the speed is independent of $\theta$, the current-induced dynamics of the Janus skyrmion is limited to a 1D motion along the interface (i.e., $\pm y$ directions) without showing the skyrmion Hall effect~\cite{Zang_PRL2011,Wanjun_NPHYS2017,Litzius_NPHYS2017}.
Both the speed and motion direction of the Janus depend on $\theta$, as shown in Figs.~\ref{FIG3}(a) and~\ref{FIG3}(b) (Supplementary Videos \blue{1} and \blue{2}~\cite{SM}). It can move toward the $+y$ or $-y$ direction, and its speed varies with $\theta$ for a given $j$.

As a larger $j$ may lead to slight deformation of the Janus skyrmion, the $|v_{y}|$-$\theta$ relation could be slightly asymmetric with respect to the $\theta$ axis corresponding to $|v_{y}|=0$ [Fig.~\ref{FIG3}(b)].
When $\theta$ is fixed at a given value, for example, $\theta=270$ degrees, the speed of the Janus skyrmion increases with increasing $j$ and a larger $j$ also leads to more obvious deformation of the Janus skyrmion [Fig.~\ref{FIG4}(a)], which can be seen from the time-dependent $m_z$ during the current-induced motion [Fig.~\ref{FIG4}(b)] (Supplementary Video \blue{3}~\cite{SM}).

A conventional skyrmion is usually repelled by magnetic surfaces and interfaces, while the Janus skyrmion is an interfacial state attracted by its hosting interface. Hence, the Janus skyrmion can be viewed as a unique topological quasiparticle strictly confined in a 1D potential well~\cite{Reichhardt_PRB2016A}.
Its current-induced dynamics could be described by the Thiele equation~\cite{Thiele_PRL1973,Zhang_NC2017,Zhang_Laminar2023,Tomasello_SREP2014} with the assumption that an interface-induced force $\boldsymbol{F}_{x}$ is acting on the Janus skyrmion and $v_{x}=0$.
The velocity solution is thus given as
$v_{y}=c\cdot\cos{(\theta+\pi/4)}$ with $c=\left|\sqrt{2}u/2\alpha\D\right|\cdot\I$ being a constant determined by the driving force $u\sim j$ and the Janus skyrmion profile (Supplementary Note \blue{1}~\cite{SM}).
Note that $\D$ is the diagonal entries of the dissipative tensor in the Thiele equation and $\I$ is related to the spin-torque efficiency over the skyrmion~\cite{Thiele_PRL1973,Zhang_NC2017,Zhang_Laminar2023,Tomasello_SREP2014}.
In Fig.~\ref{FIG3}, it can be seen that the simulation results are in a good agreement with the theoretical velocity solutions, where $c=20$ for $j=1$ MA cm$^{-2}$ and $c=60$ for $j=3$ MA cm$^{-2}$.
Note that a current-driven Janus skyrmion may detach from the interface when $\boldsymbol{F}_{x}$ is too strong (Supplementary Fig.~\blue{3}~\cite{SM}), which then transforms into a conventional skyrmion (Supplementary Videos~\blue{4} and~\blue{5}~\cite{SM}).
Besides, the imperfections at the interface may reduce the Janus skyrmion speed and induce instability, especially at high current densities (Supplementary Fig.~\blue{4}~\cite{SM}).
We also compare the current-induced motion of a Janus skyrmion with that of conventional N{\'e}el-type and Bloch-type skyrmions in Supplementary Note \blue{1}~\cite{SM}.

\begin{figure}[t]
\centerline{\includegraphics[width=0.50\textwidth]{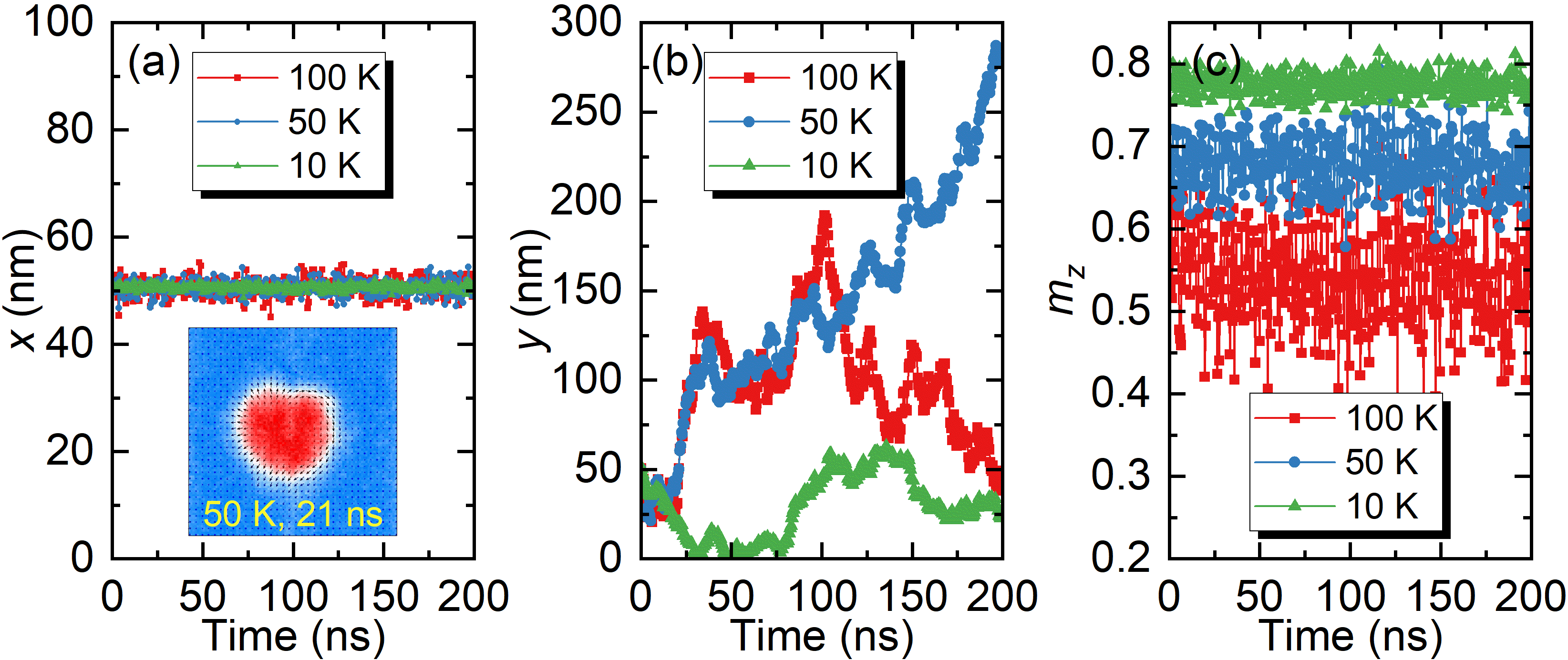}}
\caption{%
1D Brownian random walk of a Janus skyrmion at the interface.
(a) Time-dependent position of the Janus skyrmion in the $x$ dimension.
(b) Time-dependent position of the Janus skyrmion in the $y$ dimension.
(c) Time-dependent $m_z$ of the model.
Here, the Brownian dynamics of the Janus skyrmion is simulated at three different temperatures ($T=10$, $50$, and $100$ K).
See Fig.~\ref{FIG3} caption for other parameters.
}
\label{FIG5}
\end{figure}

\subsection{One-dimensional random walk}
\label{se:Random}

As the Janus skyrmion can only move along the interface, it should be able to demonstrate 1D random walk at the interface.
To examine this prediction, we introduce a thermal fluctuation term into the LLG equation~\cite{MuMax}, and simulate the Brownian motion of a Janus skyrmion at finite temperature.
The spin dynamics at finite temperature is governed by the stochastic LLG equation $\partial_{t}\boldsymbol{m}=-\gamma_{0}\boldsymbol{m}\times(\boldsymbol{h}_{\text{eff}}+\boldsymbol{h}_{\text{f}})+\alpha(\boldsymbol{m}\times\partial_{t}\boldsymbol{m})$,
where $\boldsymbol{h}_{\text{f}}$ is a thermal fluctuating field satisfying
$<h_{i}(\boldsymbol{x},t)>=0$ and
$<h_{i}(\boldsymbol{x},t)h_{k}(\boldsymbol{x}',t')>=\frac{2\alpha k_{\text{B}}T}{M_\text{S}\gamma_0\mu_0 V}\delta_{ik}\delta(\boldsymbol{x}-\boldsymbol{x}')\delta(t-t')$.
$i$ and $k$ are Cartesian components, $k_{\text{B}}$ is the Boltzmann constant, $T$ is the temperature, and $V$ is the volume of a single mesh cell.
$\delta_{ik}$ and $\delta(\dots)$ denote the Kronecker and Dirac delta symbols, respectively.
In all simulations, a fixed finite-temperature integration time step of $10$ fs is applied.
%

In Fig.~\ref{FIG5}, it can be seen that the Janus skyrmion demonstrates obvious 1D random walk along the interface (Supplementary Videos \blue{6}-\blue{8}~\cite{SM}).
The random walk diffusion in the $y$ dimension increases with $T$ [Fig.~\ref{FIG5}(b)], but the skyrmion position in the $x$ dimension only slightly fluctuates with time [Fig.~\ref{FIG5}(a)].
The Janus skyrmion size also increases with $T$, which can be seen from time-dependent $m_z$ in Fig.~\ref{FIG5}(c).
The 1D Brownian motion is a unique thermal dynamic feature of interfacial quasiparticles. A conventional skyrmion usually demonstrates 2D Brownian motion~\cite{Tretiakov_PRL2016,Miltat_PRB2018,Nozaki_APL2019,Zhao_PRL2020,Jing_PRB2021,Zhou_PRB2021,Miki_JPSJ2021,Zhang_NL2023,Zhang_ADPREPRINT2025}.

\section{Conclusion}
\label{se:Conclusion}

We have explored the unique dynamics of a Janus skyrmion induced by magnetic field, spin current, or temperature.
The Janus skyrmion is stabilized at the interface between two magnetic regions with different types of antisymmetric exchange interactions. Its helicity structure include both N{\'e}el-type and Bloch-type components.
The dynamic behaviors of a Janus skyrmion largely differ from that of a conventional skyrmion as it is topological quasiparticle stabilized and attached to the interface. The dynamic degree of freedom of a Janus skyrmion is limited by the interface, and consequently, the speed of a Janus skyrmion moving along the interface could be controlled by the spin polarization angle, and it will not show the skyrmion Hall effect.
The skyrmion Hall effect is a dynamic feature intrinsic to conventional ferromagnetic skyrmions~\cite{Reichhardt_RMP2022,Zang_PRL2011,Wanjun_NPHYS2017,Litzius_NPHYS2017,Tomasello_SREP2014,Zhang_Laminar2023,Raab_PRE2024,Zhang_PNAS2025,Dai_SA2023}, which could be avoided in antiferromagnetic systems~\cite{Xichao_NC2016,Xichao_PRB2016A,Zhang_SR2016AFM,Tretiakov_PRL2016,Pham_SCIENCE2024}.
Moreover, thermal fluctuations lead to 1D random walk of the Janus skyrmion at the interface, which is different to conventional skyrmions that can diffuse in two dimensions.
It is also found that both in-plane and out-of-plane magnetic fields can modify the size of a Janus skyrmion due to its helicity asymmetry. The size of conventional skyrmions with symmetric helicity structures cannot be adjusted by an in-plane magnetic field.
The asymmetric helicity structures of heart-shaped Janus skyrmions result in distinctive responses to magnetic field, spin current, and temperature.
As the Janus skyrmion does not show the skyrmion Hall effect, it could be a promising building block for racetrack-type device applications~\cite{Parkin_2008,Parkin_2015,Parkin_2022,Yu_NL2023,Friedman_APL2021}.
This work will pave the way for the understanding of interfacial topological quasiparticles.

\begin{acknowledgments}
X.Z. and M.M. acknowledge the support by the CREST, the Japan Science and Technology Agency (Grant No.~JPMJCR20T1).
X.Z. acknowledges the support by the Grants-in-Aid for Scientific Research from JSPS KAKENHI (Grants No.~JP25K17939 and No.~JP20F20363).
R.Z. acknowledges the support by the State Key Laboratory of Displays and Opto-Electronics (Project Reference:~ITC-PSKL12EG02).
Q.S. acknowledges the support by the RGC General Research Fund (No. 16309924).
Y.Z. acknowledges the support by the Shenzhen Fundamental Research Fund (Grant No.~JCYJ20210324120213037), the National Natural Science Foundation of China (Grant No.~12374123), the Guangdong Basic Research Center of Excellence for Aggregate Science, and the 2023 SZSTI Stable Support Scheme.
C.R. and C.J.O.R. acknowledge the support by the U.S. Department of Energy through the Los Alamos National Laboratory. Los Alamos National Laboratory is operated by Triad National Security, LLC, for the National Nuclear Security Administration of the U.S. Department of Energy (Contract No.~892333218NCA000001).
M.M. acknowledges the support by the Grants-in-Aid for Scientific Research from JSPS KAKENHI (Grants No.~JP25H00611, No.~JP24H02231, No.~JP23H04522, and No.~JP20H00337) and the Waseda University Grant for Special Research Projects (Grant No.~2025C-133).
\end{acknowledgments}



\end{document}